\documentclass[amsmath,eqsecnum,preprint,pre,aps,showpacs]{revtex4}
\usepackage{amsmath}
\usepackage{graphicx}

\begin{document}

\title{Aging of a Homogeneously Quenched Colloidal Glass-forming Liquid}
\author{ Pedro Ram\'irez-Gonz\'alez and Magdaleno Medina-Noyola}

\address{Instituto de F\'{\i}sica {\sl ``Manuel Sandoval
Vallarta"}, Universidad Aut\'{o}noma de San Luis Potos\'{\i},
\'{A}lvaro Obreg\'{o}n 64, 78000 San Luis Potos\'{\i}, SLP,
M\'{e}xico}
\date{\today}

\begin{abstract}
The non-equilibrium self-consistent generalized Langevin equation
theory of colloid dynamics is used to describe the non-stationary
aging processes occurring in a suddenly quenched model colloidal
liquid with hard-sphere plus short-ranged attractive interactions,
whose static structure factor and van Hove function evolve
irreversibly from the initial conditions before the quench to a
final, dynamically arrested state. The comparison of our numerical
results with available simulation data are highly encouraging.

\bigskip
\bigskip
\bigskip
\vskip2cm

\end{abstract}


\maketitle

\section{Introduction.}\label{secI}

The non-stationary, slowly-evolving dynamics of deeply quenched
fluids, referred to as aging, has been the subject of considerable
attention over the last decade \cite{cipelletti1,leheny1}.
Concentrated emulsions \cite{emulsions}, colloidal gels
\cite{colloidalgels}, and aqueous clay suspensions \cite{clays} are
some examples of aging systems. In spite of the apparent diversity
of these structurally-disordered and out-of-equilibrium materials,
the appearance of certain universal features in their
non-equilibrium evolution suggests the existence of an underlying
common source of the observed dynamic properties. Although this
non-stationary behavior is associated with the formation of
disordered solids, including hard materials such as polymer glasses
\cite{polymerglasses}, the main features are best exhibited by soft
materials such as those above. In particular, the study of the
dynamic properties of aging colloidal glasses and gels is specially
interesting, since the observations provided, for example, by
experimental methods such as dynamic light scattering
\cite{vanmegenmortensen,pham1,elmasriaging,martinezvanmegen} can
sometimes be complemented with direct visualizations at the level of
individual particles by means of digital video imaging techniques
\cite{sanz1,lu1}. Computer simulation experiments in well-defined
model systems have also contributed with important complementary
information about the general properties of aging
\cite{kobbarrat,foffiaging1,puertas1}.

From the theoretical side the study of aging has been addressed in
the field of spin glasses, where a mean-field theory has been
developed within the last two decades \cite{cugliandolo1}. The
models involved, however, lack a geometric structure and hence
cannot describe the spatial evolution of real colloidal glass
formers. About a decade ago Latz \cite{latz} attempted to extend the
mode coupling theory (MCT) of the ideal glass transition
\cite{goetze1,goetze2} to describe the irreversible relaxation,
including aging processes, of a suddenly quenched glass forming
system. Similarly, De Gregorio et al. \cite{degregorio} discussed
time-translational invariance and the fluctuation-dissipation
theorem in the context of the description of slow dynamics in system
out of equilibrium but close to dynamical arrest. Unfortunately, in
neither of these two theoretical efforts quantitative predictions
were presented that could be contrasted with experimental or
simulated results in specific model systems of structural
glass-formers.

For concreteness, let us focus our discussion on the conceptually
simplest glass-forming system, namely, a mono-component fluid made
of $N$ identical spherical particles in a volume $V$ which interact
through the pair potential $u(r)$ (although in the experimental
realization of this idealized model we probably have to consider a
small amount of polydispersity to suppress the kinetic pathway to
ordered phases). Assume that in the absence of external fields this
system is initially prepared in an equilibrium state corresponding
to a mean density $\overline{n}^{(0)}=N/V$ and a temperature
$T^{(0)}$, in which the static structure factor is
$S^{(0)}(k)=S^{eq}(k;n,T^{(0)})$. In the simplest idealized quench
experiment, at the time $t=0$ the temperature of the system is
instantaneously and discontinuously changed to a value $T^{(f)}$.
Let us assume that along the process that follows the quench, the
density and the temperature are constrained to remain uniform and
constant, i.e., that $\overline{n}(\textbf{r},t) =
\overline{n}^{(0)}$ and $T(\textbf{r},t) = T^{(f)}$ at any position
\textbf{r} in the volume $V$ and any time $t>0$. The relevant
question then refers to the value of the static structure factor
$S(k;t)$ for $t>0$, and to the evolution of the dynamic properties
of the system along this process.

The referred dynamic properties can be described in terms of the
relaxation of the fluctuations $\delta n({\bf r} ,t)$ of the local
concentration $n({\bf r},t)$ of colloidal particles around its bulk
equilibrium value $n$. The average decay of $\delta n({\bf r},t)$ is
described by the two-time correlation function $F(k,\tau;t)\equiv
N^{-1}\overline{ \delta n({\bf k},t+\tau)\delta n(-{\bf k},t)} $ of
the Fourier transform $\delta n({\bf k},t)$ of the fluctuations
$\delta n({\bf r} ,t)$, whose equal-time limit is $S(k;t)\equiv
F(k,\tau=0;t) = N^{-1}\overline{ \delta n({\bf k},t)\delta n(-{\bf
k},t)} $. We refer to the time $\tau$ as the \emph{correlation
time}, and the overline refers to the average over the probability
distribution of the \emph{non-equilibrium} ensemble that governs the
statistical properties of $\delta n({\bf r},t)$ at the evolution
time $t$. This ensemble will surely coincide with an equilibrium
ensemble only in the limit $t \to \infty$, provided that no dynamic
arrest condition appears along the process.

After the sudden temperature change at $t=0$ has occurred the system
evolves spontaneously, searching for its new thermodynamic
equilibrium state, at which the static structure factor should be
$S^{eq}(k;n^{(0)},T^{(f)})$. If the end state, however, is a
dynamically-arrested state (a glass or a gel), the system may never
be able to reach this equilibrium state within experimental times;
one then refers to the evolution time $t$ as the \emph{waiting} or
\emph{aging} time
\cite{pham1,cipelletti1,martinezvanmegen,sanz1,lu1}. The dependence
of $S(k;t)$ and $F(k,\tau;t)$ on $t$ characterizes the
non-equilibrium evolution of the system, whose quantitative
theoretical first-principles description, to our knowledge, has not
been available until now, in spite of important theoretical efforts
like those referred to above.

In recent related work \cite{nescgle1}, however, an extension was
proposed of the self-consistent generalized Langevin equation
(SCGLE) theory of colloid dynamics
\cite{scgle0,scgle1,scgle2,marco1,marco2} and dynamic arrest
\cite{rmf,todos1,todos2,attractive1,soft1,rigo1,rigo2,luis1}, aimed
precisely at describing this non-equilibrium evolution of $S(k;t)$
and $F(k,\tau;t)$. This extension was based on Onsager's theory of
thermal fluctuations \cite{onsager1,
onsager2,onsagermachlup1,onsagermachlup2,keizer}, adequately
extended \cite{delrio,faraday} to allow for the description of
memory effects. The purpose of the present paper is to provide the
first practical and concrete application of such general
non-equilibrium theory of colloid dynamics by means of its use in
the quantitative description of the aging process of a model
monocomponent glass-forming liquid.

In this particular context, such \emph{non-equilibrium}
self-consistent generalized Langevin equation (NE-SCGLE) theory
consists of a closed self-consistent system of equations for
$S(k;t)$ and $F(k,\tau;t)$, which we numerically solve here for a
model mono-component fluid of particles interacting through the
hard-sphere plus short-ranged attractive Yukawa potential. This
model system exhibits the glass-fluid-glass reentrance predicted by
the equilibrium SCGLE theory \cite{attractive1} (and originally
discovered by MCT \cite{bergenholtz1}). Here we discuss the
isochoric quench of this fluid from an initial equilibrium state
($n^{(0)},T^{(0)}$) in the reentrant fluid pocket of the ($n,T$)
state space, to a final temperature $T^{(f)}\ (<T^{(0)})$ in the
vicinity and below the attractive glass transition temperature
$T^{(a)}(n^{(0)})$ corresponding to the density $n^{(0)}$. This
process mimics the computer simulation aging experiment reported by
Foffi et al. \cite{foffiaging1} in a similar model system
(hard-sphere plus short-ranged square well). Here we discuss our
theoretical predictions in reference to the observed behavior in
this simulated quench experiment.

We start this discussion by summarizing in the following section the
full non-equilibrium self-consistent generalized Langevin equation
(NE-SCGLE) theory, which does not involve the restrictive assumption
of spatial homogeneity. In the same section this theory is
simplified according to the assumption that the system is
constrained to remain spatially homogeneous and isotropic. The
actual solution of the resulting equations are reported in Sec.
\ref{secIII}. The last section contains a summary of the results.

\section{non-equilibrium self-consistent generalized Langevin equation
theory}\label{secII}

The previous discussion implicitly assumes that $S(k;t)$ and
$F(k,\tau;t)$ adequately represent the structural and dynamic
properties of the quenched system along the irreversible
equilibration process. This assumption, which is thought to be
accurate in the absence of external fields, or when the effects of
these external fields are very small, is in reality a strong
simplifying assumption when the local concentration fluctuations do
not relax within experimental times, as it occurs at and near
dynamically arrested states. The general non-equilibrium SCGLE
theory proposed in ref. \cite{nescgle1}, however, does not
incorporate this simplifying assumption at the outset. Instead, it
describes the statistical properties of the instantaneous local
concentration profile $n(\textbf{r},t)$ of the colloidal liquid in
terms of the coupled time evolution equations for its mean value
$\overline{n}(\textbf{r},t)$ and for the covariance
$\sigma(\textbf{r},\textbf{r}';t)\equiv \overline{\delta n
(\textbf{r},t)\delta n (\textbf{r}',t)}$ of the fluctuations $\delta
n(\textbf{r},t) = n(\textbf{r},t)- \overline{n}(\textbf{r},t)$. In
this section we briefly review the general NE-SCGLE theory and then
particularize it to instantaneous homogeneous quench processes.

\subsection{General NE-SCGLE theory.}\label{subsecII.1}

The referred equations for $n(\textbf{r},t)$ and
$\sigma(\textbf{r},\textbf{r}';t)$ read \cite{nescgle1}
\begin{equation} \frac{\partial \overline{n}(\textbf{r},t)}{\partial
t} = D^0{\nabla} \cdot b(\textbf{r},t)\overline{n}(\textbf{r},t)
\nabla \beta\mu[{\bf r};\overline{n}(t)] \label{difeqdl}
\end{equation}
and
\begin{eqnarray}
\begin{split}
\frac{\partial \sigma(\textbf{r},\textbf{r}';t)}{\partial t} = &
D^0{\nabla} \cdot \overline{n}(\textbf{r},t) \ b(\textbf{r},t)\nabla
\int d \textbf{r}_1
\mathcal{E}[\textbf{r},\textbf{r}_1;\overline{n}(t)]
\sigma(\textbf{r}_1,\textbf{r}';t) \\ & +  D^0{\nabla}' \cdot
\overline{n}(\textbf{r}',t) \ b(\textbf{r}',t)\nabla' \int d
\textbf{r}_1 \mathcal{E}[\textbf{r}',\textbf{r}_1;\overline{n}(t)]
\sigma(\textbf{r}_1,\textbf{r};t) \\ & -2D^0{\nabla} \cdot
\overline{n}(\textbf{r},t)  \ b(\textbf{r},t)\nabla
\delta(\textbf{r}-\textbf{r}'), \label{relsigmadif2}
\end{split}
\end{eqnarray}
in which $D_0$ is the self-diffusion coefficient of the colloidal
particles in the absence of direct interactions, $\mu[{\bf r};n]$ is
the electrochemical potential at position \textbf{r} (which is a
\emph{functional} of the local concentration profile
$n(\textbf{r})$), and
$\mathcal{E}[\textbf{r},\textbf{r}';\overline{n}(t)]$ is the
functional derivative $\mathcal{E}[\textbf{r},\textbf{r}';n] \equiv
\left[ {\delta \beta\mu [{\bf r};n]}/{\delta n({\bf r}')}\right]$
evaluated at the concentration profile
$n(\textbf{r})=\overline{n}(\textbf{r},t)$. Thus, for given $D_0$
and $\mu[{\bf r};n]$, Eqs. (\ref{difeqdl}) and (\ref{relsigmadif2})
would constitute a closed system of equations for
$\overline{n}(\textbf{r},t)$ and $\sigma(\textbf{r},\textbf{r}';t)$
if it were not for the presence of the dimensionless local mobility
function $b(\textbf{r},t)$.

This mobility function $b(\textbf{r},t)$ describes the local
frictional effects of the direct (i.e., conservative) interactions
between the colloidal particles, as deviations from the value
$b(\textbf{r},t)=1$, and can be expressed in terms of the memory
function of the two-time correlation function
$C^{(2)}(\textbf{r},\textbf{r}';t,t')\equiv \overline{\delta n
(\textbf{r},t)\delta n (\textbf{r}',t')}$, which we write as
$C^{(2)}(\textbf{r},\textbf{r}+\textbf{x};t,t+\tau)\equiv
C(\textbf{x},\tau;\textbf{r},t)$, and which is the value of the van
Hove function at a spatial location \textbf{r} in the system and at
an evolution time $t$. Since the covariance is
$\sigma(\textbf{r},\textbf{r}';t) =
C^{(2)}(\textbf{r},\textbf{r}';t,t)=
C(\textbf{x},\tau=0;\textbf{r},t)$, it can also be written as
$\sigma(\textbf{x};\textbf{r},t)$. Although in the development of
the non-equilibrium SCGLE theory the assumption of absolute spatial
homogeneity and isotropy is avoided, these spatially-varying van
Hove function and covariance do depend on the location \textbf{r} in
space but are assumed to be approximately isotropic within a small
volume around \textbf{r}, so that they only depend on the magnitude
$|\textbf{x}|$ of the correlation vector \textbf{x}. Under these
conditions, the local covariance $\sigma(|\textbf{x}|;\textbf{r},t)$
can be written in terms of its Fourier transform
$\sigma(k;\textbf{r},t)$ as

\begin{equation}
\sigma(\mid\textbf{x}\mid;\textbf{r},t)= \frac{1}{(2\pi)^3}\int d^3
k e^{-i\textbf{k}\cdot \textbf{x}} \sigma(k;\textbf{r},t),
\label{ftsigma}
\end{equation}
so that Eq. (\ref{relsigmadif2}) may be re-written as
\begin{eqnarray}
\begin{split}
\frac{\partial \sigma(k;\textbf{r},t)}{\partial t} = & -2k^2 D^0
\overline{n}(\textbf{r},t) b(\textbf{r},t)
\mathcal{E}(k;\overline{n}(\textbf{r},t)) \sigma(k;\textbf{r},t)
\\ & +2k^2 D^0 \overline{n}(\textbf{r},t)\ b(\textbf{r},t), \label{relsigmadif2p}
\end{split}
\end{eqnarray}
where $\mathcal{E}(k;\overline{n}(\textbf{r},t)) \equiv
(2\pi)^{-3}\int d^3 k e^{-i\textbf{k}\cdot \textbf{x}}
\mathcal{E}[\textbf{r},\textbf{r}+\textbf{x};
\overline{n}(\textbf{r},t)]$. Similarly, the local van Hove function
$C(|\textbf{x}|,\tau;\textbf{r},t)$ can also be expressed in terms
of its spatial Fourier transform as
\begin{equation}
C(\mid\textbf{x}\mid,\tau;\textbf{r},t)= \frac{1}{(2\pi)^3}\int d^3
k e^{-i\textbf{k}\cdot \textbf{x}} C(k,\tau;\textbf{r},t).
\end{equation}
Let us notice that we can also introduce the notation
$C(k,\tau;\textbf{r},t)=\overline{n}(\textbf{r},t)F(k,\tau;\textbf{r},t)$,
with $F(k,\tau;\textbf{r},t)$ being the non-equilibrium intermediate
scattering function, whose initial value $ F(k,\tau=0;\textbf{r},t)=
S(k;\textbf{r},t)$ defines the time-evolving spatially-varying
static structure factor $S(k;\textbf{r},t)$; this more familiar
notation will be employed later on.

According to Ref. \cite{nescgle1}, the actual calculation of the
local mobility function $b(\textbf{r},t)$ requires the solution, at
each position \textbf{r} and each evolution time $t$, of a system of
equations involving the Laplace transform (LT) of $ C(k,\tau;
\textbf{r},t)$ (denoted by $\hat C(k,z; \textbf{r},t) \equiv
\int_0^\infty d \tau e ^{-z\tau}C(k,\tau; \textbf{r},t)$), as well
as the LT of its \emph{self} component $C_S(k,\tau; \textbf{r},t)$,
and of the $\tau$-dependent friction function $\Delta \zeta^* (\tau;
\textbf{r},t)$, namely,
\begin{gather}\label{fluct5}
\hat C(k,z; \textbf{r},t) = \frac{\sigma(k;
\textbf{r},t)}{z+\frac{k^2D^0 \overline{n}(\textbf{r},t)
\sigma^{-1}(k; \textbf{r},t)}{1+\lambda (k)\ \Delta \hat\zeta^*(z;
\textbf{r},t)}},
\end{gather}

\begin{gather}\label{fluct5s}
\hat C_S(k,z; \textbf{r},t) = \frac{1}{z+\frac{k^2D^0 }{1+\lambda
(k)\ \Delta \hat\zeta^*(z; \textbf{r},t)}},
\end{gather}
and
\begin{equation}
\Delta \zeta^* (\tau; \textbf{r},t)=\frac{D_0}{3\left( 2\pi \right)
^{3}}\int d {\bf k}\ k^2 \left[\frac{ \sigma(k;
\textbf{r},t)/\overline{n}(\textbf{r},t)-1}{\sigma(k;
\textbf{r},t)}\right]^2 C(k,\tau; \textbf{r},t)C_S(k,\tau;
\textbf{r},t). \label{dzdt}
\end{equation}
with $\lambda (k)$ being a phenomenological ``interpolating
function" given by \cite{nescgle1,todos1,todos2}
\begin{equation}
\lambda (k)=\frac{1}{1+\left( \frac{k}{k_{c}}\right) ^{2}},
\label{lambdadk}
\end{equation}
where $k_{c } \gtrsim 2\pi/d$, with $d$ being some form of distance
of closest approach. A simple empirical prescription is to choose
$k_{c }$ as $k_{c }= k_{\min }$, the position of the first minimum
(beyond the main peak) of the static structure factor $S(k;
\textbf{r},t)=\sigma(k; \textbf{r},t)/\overline{n}(\textbf{r},t)$.
The local mobility $b(\textbf{r},t)$ finally follows from the
solution of these equations by means of its relation with $\Delta
\hat\zeta^*(z; \textbf{r},t)$, namely,
\begin{equation}
b(\textbf{r},t)= \left[1+\Delta\hat {\zeta}^*(z=0;
\textbf{r},t)\right]^{-1}. \label{bdkz}
\end{equation}

\subsection{Instantaneous homogeneous quench.}\label{subsecII.2}

Let us now discuss the application of this general theory to the
particular conditions referring to the irreversible evolution of the
structure and dynamics of a system \emph{constrained} to suffer a
programmed process of \emph{homogeneous} compression or expansion
(and/or of cooling or heating). Under these conditions, rather than
solving Eq. (\ref{difeqdl}) for $\overline{n}({\bf r};t)$, we assume
that the system is constrained to remain \emph{spatially uniform},
$\overline{n}({\bf r};t)=\overline{n}(t)$, according to a
\emph{prescribed} time-dependence $\overline{n}(t)$ of the uniform
bulk concentration  (and/or to a prescribed uniform time-dependent
temperature $T(t)$). In consistency with this assumed constraint we
have that the dependence on the position \textbf{r} disappears from
the previous equations so that, for example, Eq.
(\ref{relsigmadif2p}) may be re-written as
\begin{equation}
\frac{\partial \sigma(k;t)}{\partial t} = -2k^2 D^0\overline{n}(t)
b(t)\mathcal{E}(k;t) \left[\sigma(k;t)
-\mathcal{E}^{-1}(k;t)\right], \label{relsigmadif2pp}
\end{equation}
with $\mathcal{E}(k;t)\equiv \mathcal{E}(k;\overline{n}(t))$ and
with
\begin{equation}
b(t)= \left[1+\int_0^{\infty}d\tau\Delta{\zeta}^*(\tau;
t)\right]^{-1}, \label{bdkt}
\end{equation}
where $\Delta{\zeta}^*(\tau; t)$ is provided by the solution of the
self-consistent system in Eqs. (\ref{fluct5})-(\ref{dzdt}) for the
uniform bulk concentration $\overline{n}({\bf
r};t)=\overline{n}(t)$.

Among the many possible programmed protocols ($\overline{n}(t)$,
$T(t)$) that one could devise to drive or to prepare the system, the
simplest corresponds to the idealized quasi-static process, in which
the relaxation rate $\partial\sigma(k;t)/\partial t$ is virtually
negligible due to a virtually instantaneous ``thermalization" of
$\sigma(k;t)$ to its local equilibrium value
$\sigma^{l.e.}(k;t)\equiv 1/\mathcal{E}(k;t)$
\cite{nescgle1,keizer}. A quasi-static process, however, is a rather
unrealistic concept, at least in the limit of small wave-vectors, in
which the relaxation times diverge as $k^{-2}$, as seen in the
example below. In contrast, a far more interesting and fundamental
protocol corresponds to the opposite limit, in which the system,
initially at an equilibrium state determined by initial values of
the control parameters, $(\overline{n}^{(0)},T^{(0)})$, must adjust
itself in response to a sudden and instantaneous change of these
control parameters to new values $(\overline{n}^{(f)},T^{(f)})$,
according to the ``program" $\overline{n}(t) =
\overline{n}^{(0)}\theta (-t)+\overline{n}^{(f)}\theta (t)$ and
$T(t) = T^{(0)}\theta (-t)+T^{(f)}\theta (t)$, with $\theta (t)$
being Heavyside's step function.

Under these conditions the formal solution of Eq.
(\ref{relsigmadif2pp}) can be written, for $t > 0$, as
\begin{equation}
\sigma(k;t)=\sigma^0(k)e^{-\alpha (k)u(t)}+
[\mathcal{E}^{(f)}(k)]^{-1}\left(1-e^{-\alpha (k) u(t)}\right),
\label{solsigmadkt}
\end{equation}
where $\mathcal{E}^{(f)}(k)=\mathcal{E}(k;\overline{n}^{(f)},
T^{(f)})$ is the Fourier transform of
$\mathcal{E}[\mid\textbf{r}-\textbf{r}'\mid;\overline{n}^{(f)},T^{(f)}]
\equiv \left[ {\delta \beta\mu [{\bf r};n]}/{\delta n({\bf
r}')}\right]_{n=\overline{n}^{(f)}, T=T^{(f)}}$,
\begin{equation}
u(t) \equiv \int_0^t b(t')dt', \label{udt}
\end{equation}
and
\begin{equation}
\alpha (k) \equiv 2k^2D^0\bar n^{(f)}\mathcal{E}^{(f)}(k).
\label{alphadk}
\end{equation}
Clearly, the presence of the time-dependent mobility $b(t)$ couples
this formal solution with the self-consistent system in Eqs.
(\ref{fluct5})-(\ref{dzdt}). For the present conditions, and in
terms of the non-stationary static structure factor $S(k;t)\equiv
\sigma(k;t)/ \overline{n}^{(f)}$ and intermediate scattering
function $F(k,\tau; t)\equiv C(k,\tau; t)/ \overline{n}^{(f)}$, we
may rewrite such self-consistent system of equations as

\begin{gather}\label{fluctquench}
\hat F(k,z; t) = \frac{S(k; t)}{z+\frac{k^2D^0 S^{-1}(k;
t)}{1+\lambda (k)\ \Delta \hat\zeta^*(z; t)}},
\end{gather}

\begin{gather}\label{fluctsquench}
\hat F_S(k,z; t) = \frac{1}{z+\frac{k^2D^0 }{1+\lambda (k)\ \Delta
\hat\zeta^*(z; t)}},
\end{gather}
and

\begin{equation}
\Delta \zeta^* (\tau; t)=\frac{D_0}{3\left( 2\pi \right)
^{3}\overline{n}^{(f)}}\int d {\bf k}\ k^2 \left[\frac{ S(k;
t)-1}{S(k; t)}\right]^2 F(k,\tau; t)F_S(k,\tau; t).
\label{dzdtquench}
\end{equation}
Eqs. (\ref{bdkt})-(\ref{dzdtquench}) constitute our general
self-consistent description of the spontaneous evolution of the
structure and dynamics of an \emph{instantaneously} and
\emph{homogeneously} quenched liquid.

Of course, one important aspect of this analysis refers to the
possibility that the end state of the quench process happens to be
in the region of dynamically arrested states. For the discussion of
this important aspect it is useful to consider the long-$\tau$ (or
small $z$) asymptotic stationary solutions of Eqs.
(\ref{fluctquench})-(\ref{dzdtquench}) above. Just like in the
equilibrium SCGLE theory \cite{todos1}, these may be analyzed in
terms of the asymptotic values of these dynamic properties (the
so-called non-ergodicity parameters),  given by \cite{nescgle1}
\begin{equation}
f(k;t)\equiv \lim_{\tau\to\infty} \frac{F(k,\tau;t)}{S(k)} = \frac
{\lambda(k;t)S(k;t)}{\lambda(k;t)S(k;t)+k^2\gamma(t)} \label{fdkinf}
\end{equation}
and
\begin{equation}
f_S(k;t)\equiv \lim_{\tau\to\infty} F_S(k,\tau;t) = \frac
{\lambda(k;t)}{\lambda(k;t)+k^2\gamma(t)}, \label{fdksinf}
\end{equation}
where the squared localization length $\gamma (t)$ is the solution
of
\begin{equation}
\frac{1}{\gamma(t)} =
\frac{1}{6\pi^{2}\overline{n}^{(f)}}\int_{0}^{\infty }
dkk^4\frac{\left[S(k;t)-1\right] ^{2}\lambda^2 (k;t)}{\left[\lambda
(k;t)S(k;t) + k^2\gamma(t)\right]\left[\lambda (k;t) +
k^2\gamma(t)\right]}. \label{nep5pp}
\end{equation}
These equations are the non-equilibrium extension of the
corresponding results of the equilibrium SCGLE theory, and their
derivation  from Eqs. (\ref{fluctquench})-(\ref{dzdtquench}) follows
the same arguments as in the equilibrium case \cite{scgle2}.

In the following section we numerically solve Eqs.
(\ref{bdkt})-(\ref{dzdtquench}) for still more specific conditions,
namely, for an \emph{isochoric} quench of a model colloidal system,
in which $\overline{n}^{(f)} = \overline{n}^{(0)}$ and Eq.
(\ref{solsigmadkt}) can be written in terms of the time-evolving
static structure factor $S(k;t)=\sigma(k;t)/\overline{n}^{(f)}$ as
\begin{equation}
S(k;t)=S^0(k)e^{-\alpha (k)u(t)}+ S^{eq}_f(k) \left(1-e^{-\alpha (k)
u(t)}\right), \label{solsdkt}
\end{equation}
with $S^{eq}_f(k)\equiv
[\overline{n}^{(f)}\mathcal{E}^{(f)}(k)]^{-1}$. Let us notice that
in the limit in which the friction function $\Delta{\zeta}^*(\tau;
t)$ vanishes,  $b(t)=1$ and hence $u(t)=t$, so that Eq.
(\ref{solsdkt}) reads
\begin{equation}
S^*(k;t)=S^0(k)e^{-\alpha (k)t}+ S^{eq}_f(k) \left(1-e^{-\alpha (k)
t}\right). \label{solsdktexp}
\end{equation}
This limiting expression describes an exponential interpolation of
$S(k;t)$ between its initial value $S^0(k)$ and its final
equilibrium value $S^{eq}_f(k)\equiv
[\overline{n}^{(f)}\mathcal{E}^{(f)}(k)]^{-1}$. It is then important
to notice that the general solution $S(k;t)$ in Eq. (\ref{solsdkt})
can be written in terms of this particular solution as
$S(k;t)=S^*(k;u(t))$, with $u(t)$ given by Eq. (\ref{udt}). This
means that a sequence of static structure factors $S^*(k;u_n)$
generated by this simple exponential interpolating formula when the
time $t$ is given a sequence of values $u_n$, say $u_n=n\Delta u$
(with $n=0,1,2,...$), will be identical to the sequence  $S(k;t_n)$
generated when the exact solution in Eq. (\ref{solsdkt}) is
evaluated at a different sequence $t_n$ ($n=0,1,2,...)$, provided
that the times $u_n$ and $t_n$ are related by $u_n=\int_0^{t_n}
b(t')dt'$. This observation greatly simplifies the mathematical
analysis and the numerical method of solution of the full
self-consistent theory under the particular conditions considered
here.

The solution ${\gamma(t)}$ of  Eq. (\ref{nep5pp}) provides a dynamic
order parameter in the sense that when it is infinite we can say
that at that waiting time $t$ the system remains ergodic, whereas if
it is finite, we say that the system became dynamically arrested. A
practical manner to use this criterion is to first construct a
sequence of static structure factors $S^*(k;u_n)$ using Eq.
(\ref{solsdktexp}) for the uniform sequence $u_n=n\Delta u$ (with
$n=0,1,2,...$). Each member of this sequence is then employed as the
static input to solve self-consistently Eqs.
(\ref{fluctquench})-(\ref{dzdtquench}), thus evaluating, using Eq.
(\ref{bdkt}), a mobility sequence $b(u_n)$. Since the sequence
$S^*(k;u_n)$ is identical to the sequence $S(k;t_n)$ provided that
$u_n=\int_0^{t_n} b(t')dt'$, the mobility $b(u_n)$ must be identical
to $b(t_n)$, and the corresponding time-sequence $t_n$ can be
determined by means of the approximate recursive relation
$t_{n+1}=t_n + (\Delta u)/ b(t_n)$. If the dynamic arrest condition
occurs along this process, i.e., if a value $u^{(a)}$ exists such
that $\gamma(u)$ (determined using $S^*(k;u)$ in Eq. (\ref{nep5pp}))
is infinite for $u < u^{(a)}$ and finite for $u> u^{(a)}$, then
$b(u)\to 0$ when $u \to u^{(a)}$ from below, and it is then not
difficult to realize that the corresponding dynamic arrest time
$t^{(a)}$ will diverge and $u^{(a)}=\int_0^{\infty} b(t')dt'$. The
following numerical results illustrate the physical implications of
this singular behavior.

\section{Illustrative application.}\label{secIII}

\begin{figure}
\begin{center}
\includegraphics[scale=.28]{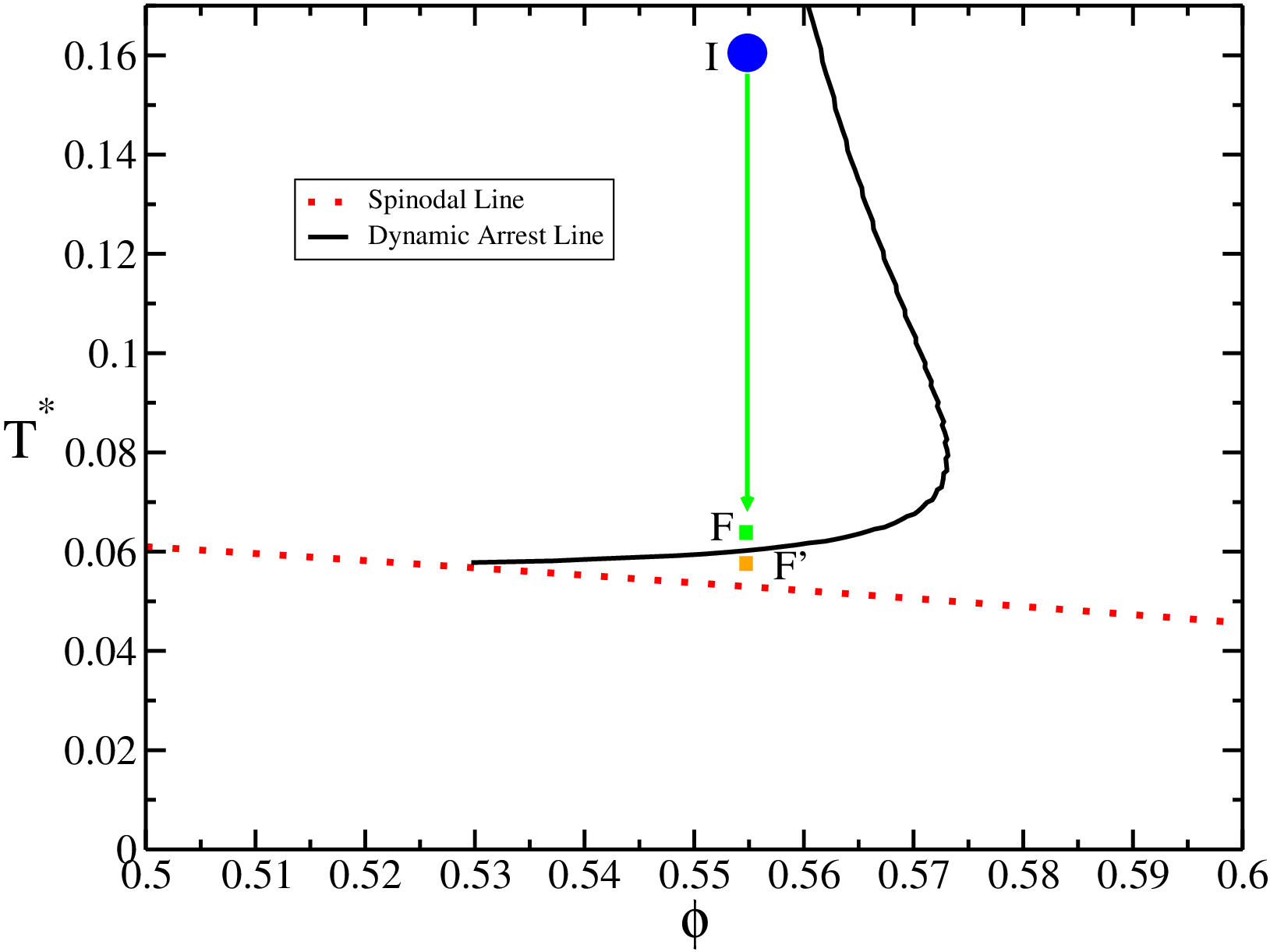}
\caption{State space $(\phi,T^*)$ of the hard-sphere plus attractive
Yukawa model system ($z=20$). The dotted line is the spinodal curve
and the solid line is the dynamic arrest line calculated using Eq.
(\ref{nep5pp}) within the mean spherical approximation (MSA) for the
equilibrium static structure factor $S^{eq}(k;\phi,T^*)$. We
consider an instantaneous quench process at $t=0$ from the ergodic
initial state $I$ to the final state $F$ near but slightly above the
attractive glass ``branch" of the dynamic arrest line. We also
consider a second process, now to the point $F'$ below this dynamic
arrest line but still above the spinodal curve.} \label{fig1}
\end{center}
\end{figure}

Let us now apply the theory just presented, to a concrete model
system, namely, a dispersion of colloidal particles interacting
through the hard-sphere plus attractive Yukawa pair potential
expressed, in units of the thermal energy $k_BT=\beta^{-1}$, as

\begin{equation}\label{yuk}
\beta u(r)=
\begin{cases}
\infty, & r < \sigma_{HS}; \\
-K\frac{\exp[-z(r/\sigma_{HS} -1)]}{(r/\sigma_{HS})}, &
r>\sigma_{HS}.
\end{cases}
\end{equation}
The state space of this system is spanned by the volume fraction
$\phi = \pi \overline{n} \sigma_{HS}^3/6$ and the reduced
temperature $T^*\equiv K^{-1}$, as illustrated in Fig. \ref{fig1}.
The equilibrium phase diagram of this system includes the gas and
liquid disordered phases and crystalline solid phases. Here we will
describe the equilibrium static structure factor
$S^{eq}(k;\phi,T^*)= [ \overline{n}
\mathcal{E}^{eq}(k;\overline{n},T^*)]^{-1}$ of the disordered phases
within the mean spherical approximation (MSA) \cite{hoyeblum}. Using
this approximation and the compressibility equation \cite{mcquarrie}
one can determine the spinodal curve of the gas-liquid transition by
means of the condition $1/S^{eq}(k=0;\phi,T^*)=0$; the result is
plotted in Fig. \ref{fig1} for $z=20$.

Using the same MSA equilibrium static structure factor
$S^{eq}(k;\phi,T^*)$ in the equilibrium version of Eq.
(\ref{nep5pp}) we can scan the state space $(\phi,T^*)$ to determine
$\gamma^{eq}$ at any point $(\phi,T^*)$ \cite{attractive1}. In this
manner one locates the dynamic arrest transition line indicated by
the solid curve of Fig. \ref{fig1}. The region to the right and
below this curve is thus predicted to correspond to dynamically
arrested states. This figure focusses on the high-density
glass-fluid-glass reentrance region that was first discovered using
mode coupling theory \cite{bergenholtz1}. We now follow the approach
introduced by Foffi et al. \cite{foffiaging1} in a simulation
experiment on a very similar model system (a hard-sphere plus
square-well fluid). Such experiment corresponds to suddenly
quenching the system under isochoric conditions from an initial
state $(\phi_0,T^*_I)$ located in the fluid pocket of the reentrance
(point $I$ in Fig. \ref{fig1}), to a final state near the
fluid-``attractive glass" transition line (either point $F$ or point
$F'$ in Fig \ref{fig1}). In the first case the end state
$(\phi_0,T^*_F)$  lies slightly above the transition line, whereas
in the second, the end state $(\phi_0,T^*_{F'})$ lies in the region
of arrested states.

\begin{figure}
\begin{center}
\includegraphics[scale=.28]{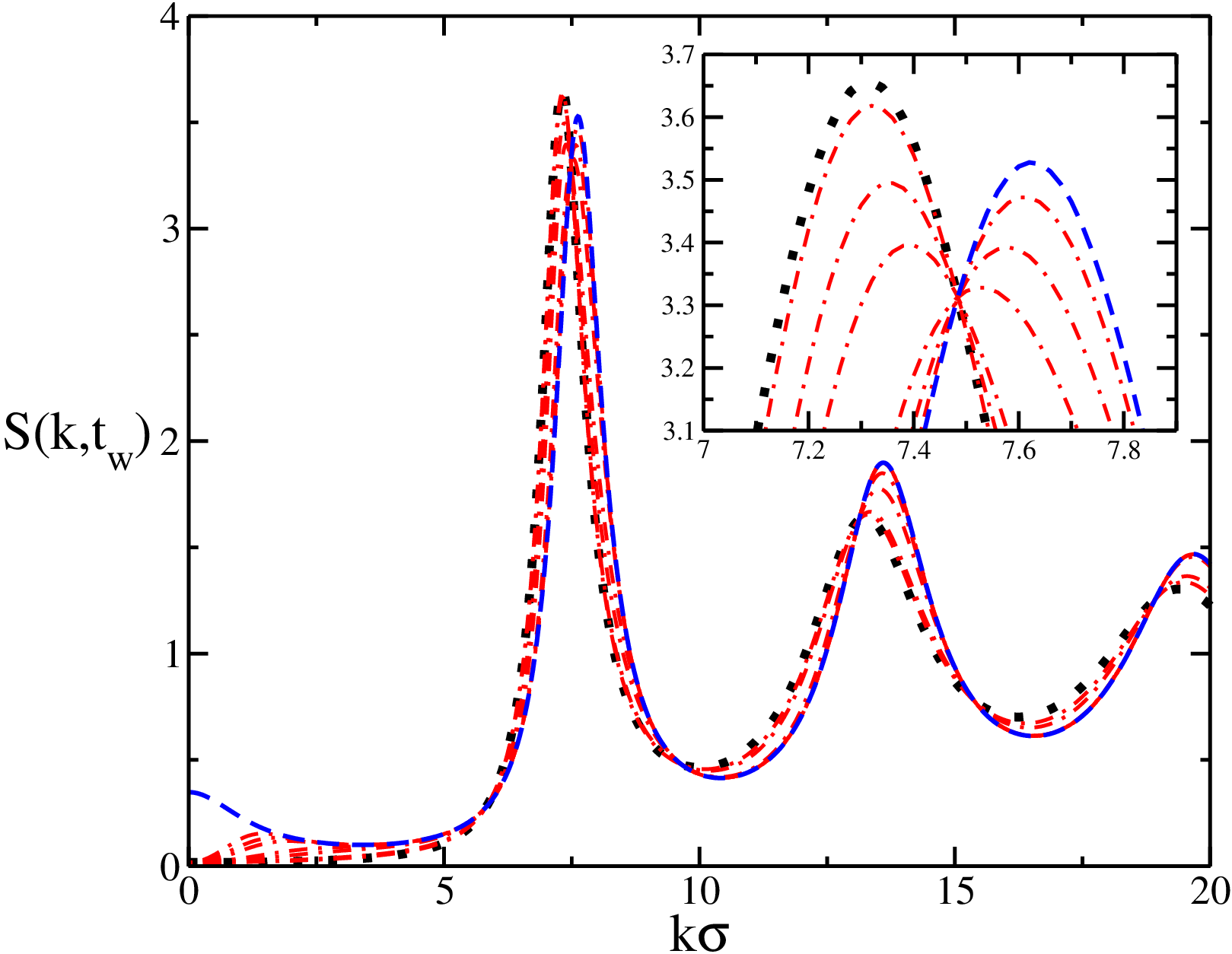}
\caption{Non-equilibrium evolution of the static structure factor
$S(k,t_{w})$. The system, initially equilibrated at
$(\phi_0,T_I^{*})=(0.555,0.159)$, with
$S(k;t=0)=S^{eq}(k;\phi_0,T_I^*)$ ((black) dotted curve), is
instantaneously quenched at $t_w=0$ to the final point
$(\phi_0,T_F^{*})=(0.555,0.0604)$. The static structure factor then
evolves continuously along a sequence of non-equilibrium values
((red) point-dashed lines) illustrated by the snapshots
corresponding to $t_{w}/t_0 = 3.2,\ 60.87,\ 174.29,\ 945.39, \
2023.54,\  \textrm{and}\ 4858.84$, with $ t_0 \equiv
[\sigma^2/D_0]$. Since the point $F$ lies outside the dynamic arrest
region, $S(k,t_{w})$ eventually attains its final equilibrium value
$S(k;t_{w}=\infty)=S^{eq}(k;\phi_0,T_F^*)$ ((blue) dashed curve).
The main figure shows the resulting relaxation process in a wide
$k$-range and the inset zooms on the evolution of the main peak of
$S(k,t_{w})$. } \label{fig2}
\end{center}
\end{figure}

For this process we solve the general self-consistent system of
equations in Eqs. (\ref{bdkt})-(\ref{dzdtquench}). The specific
calculations are performed along the isochore $\phi_0=0.555$ with
initial temperature $T_I^{*}=0.159$ and final temperature
$T^*_F=0.0604$.  Fig. \ref{fig2} illustrates the irreversible
evolution of the static structure factor $S(k;t_w)$ as a sequence of
snapshots corresponding to five intermediate waiting times $t_w$
(from now on denoted by $t_w$, rather than simply by $t$). We
observe that the structure, initially described by
$S^{eq}(k;\phi_0,T_I^*)$, relaxes to the expected final value
$S^{eq}(k;\phi_0,T_F^*)$, and that this process is faster at large
wave-vectors, where it involves the appearance of stronger
oscillations with $k$ and a general shift of the maxima of $S(k;t)$
to larger wave-vectors. To a large extent these features can be
understood in terms even of the simple interpolating expression in
Eq. (\ref{solsdktexp}).

The corresponding adjustment of the main peak of $S(k;t_{w})$ from
its initial value $S^{eq}(k_{max};\phi_0,T^*_I)$ to its final value
$S^{eq}(k_{max};\phi_0,T^*_F)\ (< S^{eq}(k_{max};\phi_0,T^*_I))$
occurs, however, notoriously more slowly than at large wave-vectors
and in an apparently non-monotonic manner, as illustrated in the
inset of Fig. \ref{fig2}, which zooms on the evolution of the main
peak. As observed there, as the system evolves, the maximum of
$S(k;t_{w})$ moves to the right while decreasing in height to a
value smaller than $S^{eq}(k_{max};\phi_0,T^*_F)$, bouncing back at
later times to reach this final value. The origin of the predicted
non-monotonic behavior can also be understood on the basis of the
simple interpolation expression in Eq. (\ref{solsdktexp}), which
implies that $S(k;t)$ will not change with waiting time for the
wave-vectors $k^*$ at which the initial and the final static
structure factors are already identical,
$S^{eq}(k^*;\phi_0,T^*_I)=S^{eq}(k^*;\phi_0,T^*_F)$. It is then not
difficult to see that if the condition
$k^{(I)}_{max}<k^*<k^{(F)}_{max}$ occurs, as it happens in our
example, we shall observe this non-monotonic effect.

\begin{figure}
\begin{center}
\includegraphics[scale=.28]{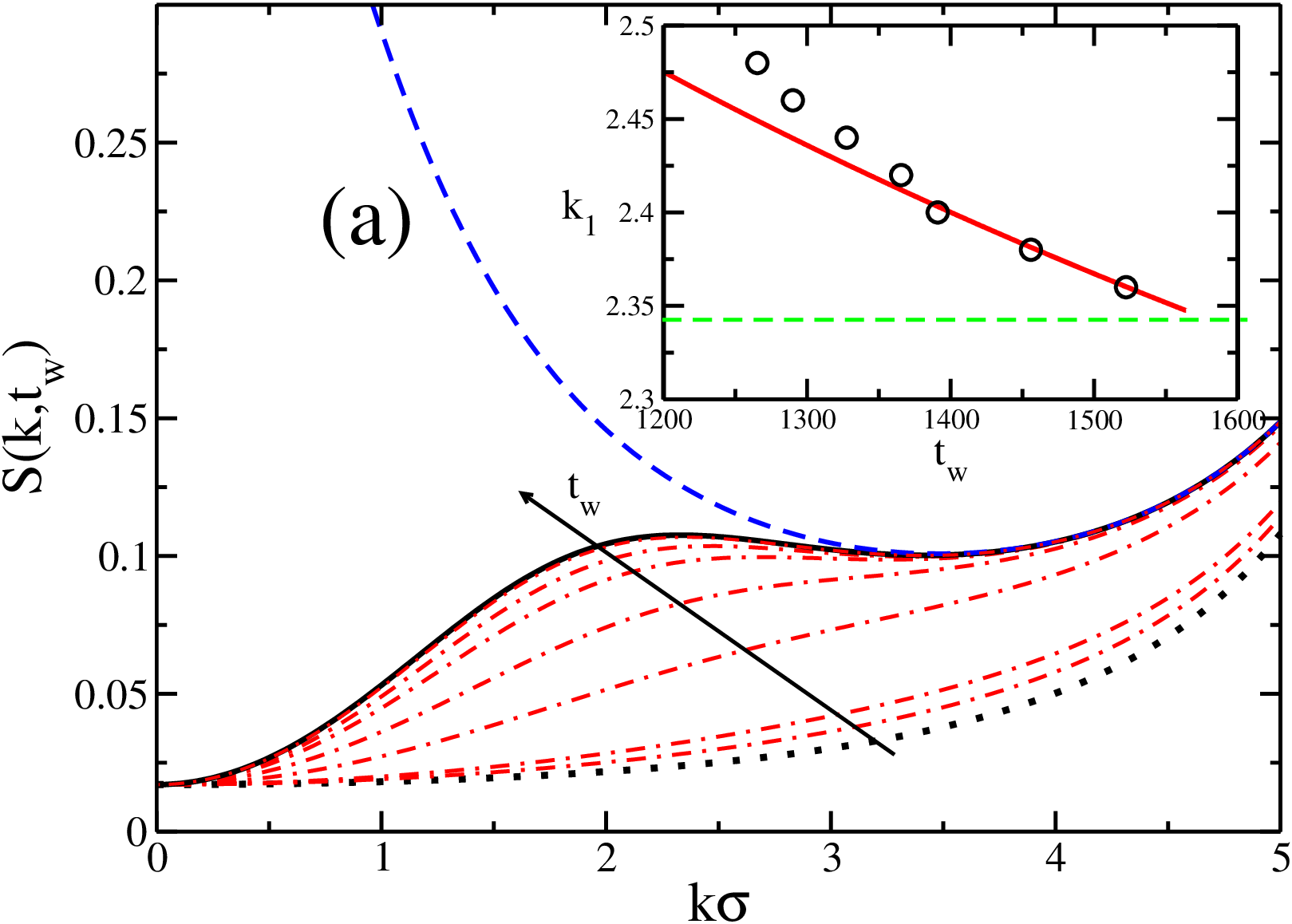}\hfill
\includegraphics[scale=.28]{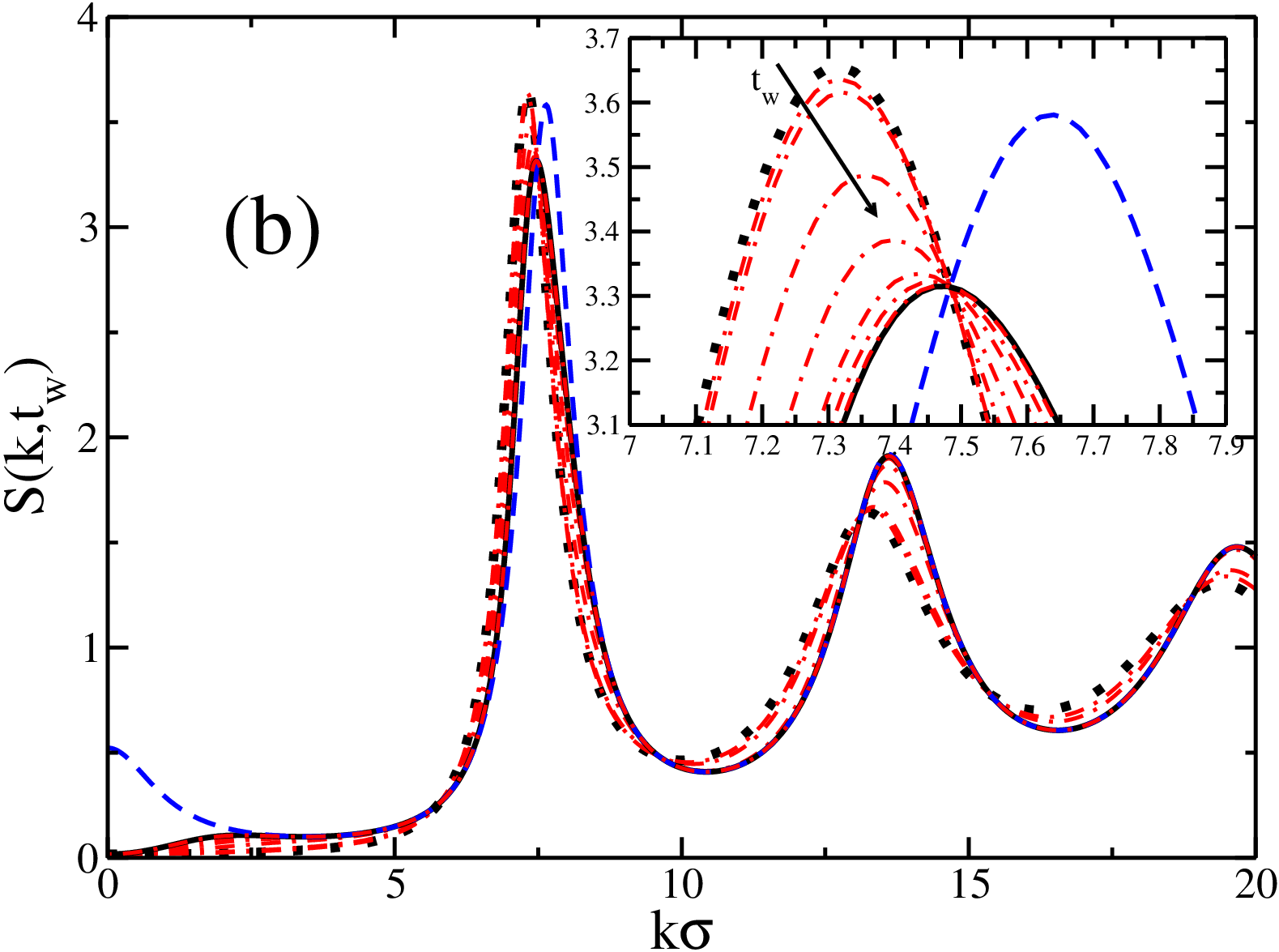}
\caption{Non-equilibrium evolution of the static structure factor
$S(k,t_{w})$ for the deeper quench to the final state point $F'$.
The system, initially equilibrated at
$(\phi_0,T_I^{*})=(0.555,0.159)$, with
$S(k;t=0)=S^{eq}(k;\phi_0,T_I^*)$ ((black) dotted curve), is
instantaneously quenched at $t_w=0$ to the final
point$(\phi,T_{F'})=(0.555,0.0588)$ inside the dynamic arrest
region. The static structure factor then evolves continuously along
a sequence of non-equilibrium values ((red) point-dashed lines)
illustrated by the snapshots corresponding to $t_{w}/t_0 =0.0,\
1.23,\ 3.84,\ 140,\ 490,\ 975,\ 1265,$ and $1522$ ($\approx
t_{w}^{(c)}/ t_0 $). Panel (a) focuses on the low-$k$ peak of
$S(k,t_w)$, and its inset shows the dependence of the position
$k_1(t_w)$ of this low-$k$ peak on the waiting time $t_w$ (empty
circles), with the solid line being the fit of the last few points
with $k_1(t_{w}) \approx (t_{w})^{-\alpha}$ and $\alpha \approx
\frac{1}{5}$. Panel (b) shows the behavior in a larger $k$-regime,
similar to Fig. \ref{fig2}, with its inset zooming on the evolution
of the main peak.} \label{fig3}
\end{center}
\end{figure}

A more interesting effect, which is perceptible in Fig. \ref{fig2},
but which is illustrated in more detail in Fig \ref{fig3}, is the
evolution of $S(k;t_{w})$ at smaller wave-vectors. This refers to
the emergence of a non-equilibrium low-$k$ peak that indicates the
appearance of spatial heterogeneities of average size $\lambda_1
(t_{w})\approx 2\pi/k_1(t_{w})$, with $k_1$ being the position of
this emerging low-$k$ maximum.  Fig \ref{fig3}.a provides a zoom on
this effect in the case of the slightly deeper quench, now to the
final state point $F'$ in Fig. \ref{fig1}, with temperature
$T^*_{F'}=0.0588$ slightly below the dynamic arrest line. These
heterogeneities may be associated with the appearance of voids whose
average size and importance increase with waiting time, as suggested
by the increasing height of the peak and by its shift to smaller
wave-vectors observed as the system evolves. The emergence of this
peak is associated with the vicinity of the gas-liquid spinodal
region. In fact, it has the same origin as the low-$k$ peak that
characterizes the process of early spinodal decomposition
\cite{cook}, even though in our case the final state
$(\phi_0,T^*_{F'})$ lies outside the spinodal region.

As said above, this phenomenon is already observed in the shallower
quench of  Fig. \ref{fig2}. In that case, however, although the
system slows down considerably, the final structure of the
irreversible evolution of $S(k;t_{w})$ is still the expected final
equilibrium static structure factor $S^{eq}(k;\phi_0,T_F)$, i.e.,
$\lim_{t_w \to \infty} S(k;t_w) = S^{eq}(k;\phi_0,T_F)$ and the
position $k_1(t_{w})$ of this low-$k$ peak decreases indefinitely.
In contrast with that scenario, in the deeper quench illustrated in
Fig \ref{fig3}, the final structure of the system is no longer
$S^{eq}(k;\phi_0,T_{F'})$; instead, the asymptotic long-$t_w$ limit
of $ S(k;t_w)$ is given by $S^{(a)}(k)\equiv  S^*(k;u^{(a)})$, where
$u^{(a)}$ is the value of $u$ at which  the dynamic arrest condition
is satisfied. This value is determined using the structure factor
$S^*(k;u)$ of Eq. (\ref{solsdktexp}) as the structural input in Eq.
(\ref{nep5pp}), as discussed at the end of Sec. \ref{secII}. In Fig.
\ref{fig3}.a we can compare the non-equilibrium arrested structure
factor $S^{(a)}(k)$ with the equilibrium structure
$S^{eq}(k;\phi_0,T_F)$ that would have been attained if no dynamic
arrest condition had appeared along the equilibration process of
$S(k;t_w)$.

In the same figure we also illustrate the evolution of $S(k;t_w)$
towards its asymptotic limit $S^{(a)}(k)$ with a series of snapshots
corresponding to a set of increasing waiting times. The most
interesting feature revealed by these snapshots is the existence of
an early evolution regime, in which $S(k;t_w)$ evolves rather
quickly towards the close neighborhood of $S^{(a)}(k)$. As
illustrated by these snapshots, this occurs within a finite waiting
time $t_w^{(c)} \approx 1500 t_0$. This early regime is followed by
an asymptotic long-$t_w$ regime, in which the evolution of
$S(k;t_w)$ to actually reach the exact asymptotic value $S^{(a)}(k)$
becomes extremely slow and completely imperceptible in the scale of
the figure.

This is illustrated in the inset of Fig. \ref{fig3}(a), where we
plot the evolution of the position $k_1(t_{w})$ of the low-$k$ peak
of $S(k;t_w)$ for various waiting times between the last two
snapshots of the main figure (i.e., $1265 t_0 \le t_w \le 1522t_0$).
We notice that in this regime the last few data for $k_1(t_{w})$ may
be fitted approximately by a power law $k_1(t_{w})  \approx 10.22
\times (t_{w})^{-1/5}$. In fact, the crossover waiting time
$t_w^{(c)}$ can be estimated more accurately by the condition $10.22
\times (t_{w}^{(c)})^{-1/5} = k_1^{(a)}$, with $k_1^{(a)}=2.34$
being the asymptotic value of $k_1(t_{w})$ corresponding to
$S^{(a)}(k)$. This yields $t_{w}^{(c)} \approx 1589t_0$. The slow
evolution regime $t_{w} > t_{w}^{(c)}$, corresponding to
asymptotically long times, cannot be observed, by definition, in the
structural evolution illustrated in Fig. \ref{fig3}. It can,
however, be observed in the evolution of the dynamic properties, as
we discuss below.

Let us emphasize the difference between the two quench processes
just discussed (i.e., those involving the final state $F$ or $F'$).
For this, Fig. \ref{fig3}.b plots the evolution of $S(k;t_{w})$ for
the latter process in the same manner as Fig. \ref{fig2} does for
the former. Let us point out that the quench simulated by Foffi et
al. \cite{foffiaging1} corresponds to the conditions illustrated in
Fig. \ref{fig3}, i.e., to the process ending in the state $F'$ just
below the dynamic arrest line. We recall that in the process
illustrated in Fig. \ref{fig2} nothing prevents the evolution of
$S(k;t_{w})$ from reaching the final structure factor
$S^{eq}(k;\phi_0,T_F)$, and this leads to the upturn of the peak
illustrated in the inset of that figure. In contrast, as observed in
the inset of Fig. \ref{fig3}.b, the main difference is that now the
main peak of $S(k;t_{w})$ decreases but seems to stop evolving when
$t_{w}$ reaches $t_{w}^{(c)}$, and this happens to occur before the
upturn of the peak towards $S^{eq}(k;\phi_0,T_{F'})$ has a chance to
develop. This is in agreement with what is observed in the simulated
quench of Foffi et al., in which the main peak only decreases
without exhibiting any upturn. On the other hand, our results in the
inset of Fig. \ref{fig3}.b also predict that the peak shifts
slightly to the right, but in the simulation results such a shift is
not appreciable.

Similarly, in the report of the simulated quench of Foffi et al.
\cite{foffiaging1} no reference is made to the low-$k$ peak
predicted by our theory according to the illustrative results in Fig
\ref{fig3}.a. Thus, at this stage we cannot make a definitive
statement on the level of a fine quantitative comparison between our
theoretical predictions and the simulation results for the evolution
of $S(k;t_{w})$, partially because of the differences in the model
and in the conditions (volume fraction, for example) in which the
quench was performed. While it is clearly desirable to carry out a
systematic comparison on identical conditions, the agreement with
important features observed in the simulation experiments is
encouraging.

\begin{figure}
\begin{center}
\includegraphics[scale=.28]{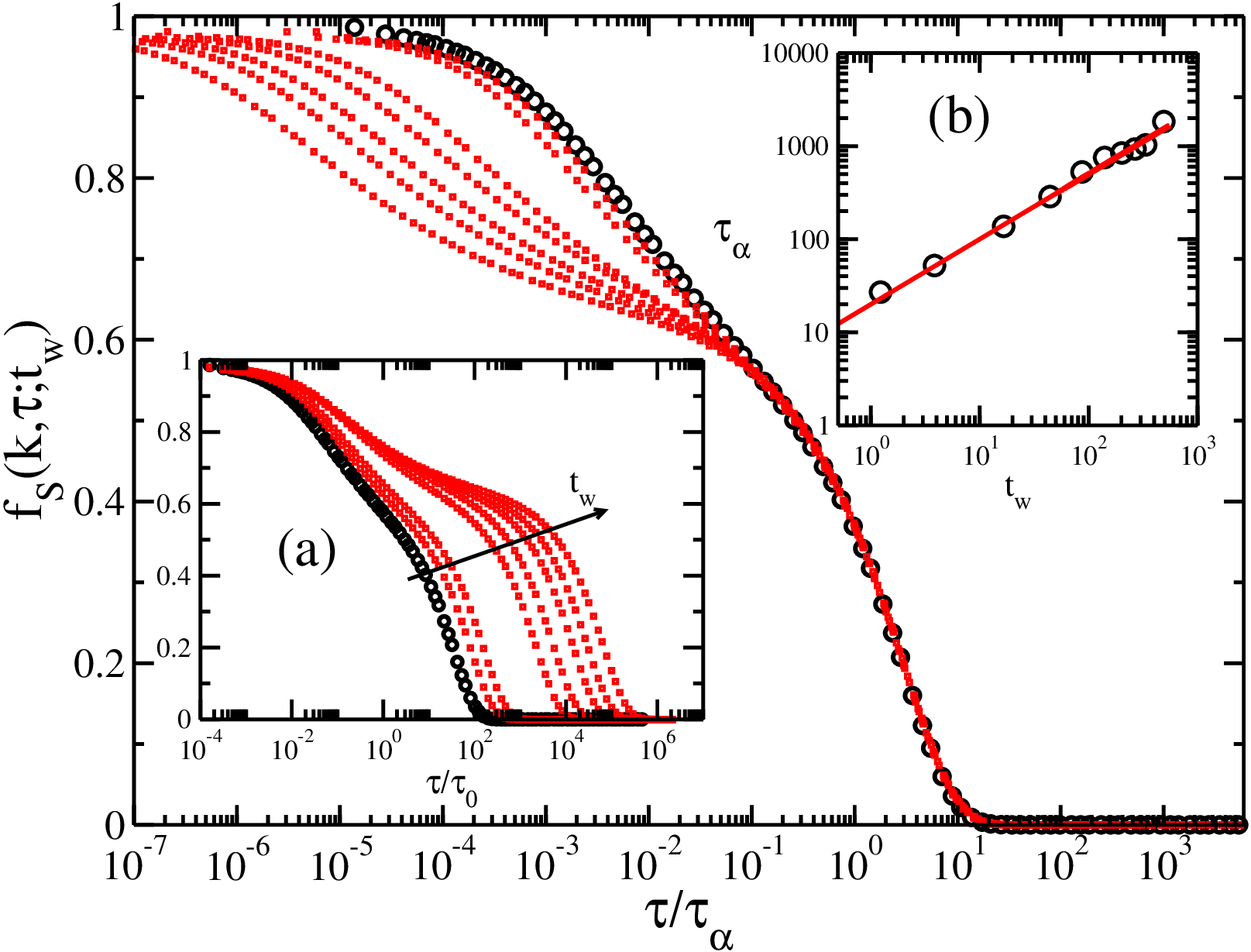}
\caption{Theoretical predictions for the dependence of the
intermediate scattering function $F(k,\tau;t_{w})$ on correlation
time $\tau$ for the quench to the final state $F'$ corresponding to
the waiting times  $t_{w}/t_0 =0.0,\ 1.23,\ 3.84,\ 140,\ 490,\ 975,\
1265,$ and $1522$ ($\approx t_{w}^{(c)}/ t_0 $). In
inset (a) and in the main figure the correlation time is scaled,
respectively, by $t_0$ and by the relaxation time $\tau_\alpha$ of
the stretched exponential fit of the final relaxation of $F(k,\tau;t_{w})$.
Inset (b) plots $\tau_\alpha$ as a function of waiting time, with the line
being the fit $\tau_\alpha \approx t_w^{0.7}$ (both times in units
of $t_0$).} \label{fig4}
\end{center}
\end{figure}

Let us conclude this exercise by showing the irreversible evolution
of the $\tau$-dependence of the intermediate scattering function
$F(k,\tau; t_{w})$ for the quenching process $I\to F'$. This is
presented in inset (a) of Fig. \ref{fig4}, where the correlator
$f(k,\tau;t_{w})\equiv F(k,\tau;t_{w})/S(k;t_{w})$  is plotted as a
function of the correlation time $\tau$ at representative waiting
times corresponding to the snapshots of $S(k;t_{w})$ of Fig.
\ref{fig3}, namely,  $t_{w}/t_0 =0.0,\ 1.23,\ 3.84,\ 140,\ 490,\ 975,\
1265,$ and $1522$ ($\approx t_{w}^{(c)}/ t_0 $).
These results illustrate the fact that the decay of the temporal
correlation of the fluctuations slows down notoriously as the system
ages, developing a two-step relaxation: the initial $\beta$-relaxation
to an increasingly better defined plateau, followed by the
$\alpha$-relaxation from this plateau. This is a typical behavior
observed in the simulation and experimental studies of aging
\cite{pham1,cipelletti1,martinezvanmegen,sanz1,lu1,kobbarrat,foffiaging1,puertas1}.
Another feature associated with aging is the superposition of the
alpha relaxation at different waiting times on a single master
curve, well-fitted by a stretched exponential function
$f(k,\tau;t_w)\approx A(k;t_w) \exp [-(\tau/\tau_\alpha)^\beta]$.
Our theoretical results also exhibit this scaling property, as
demonstrated in the main panel of  Fig. \ref{fig4}. The exponent
$\beta$ is independent of $t_w$ (although it may depend on $k$). For
the case illustrated in the figure we find $\beta \approx 0.9$. The
$\alpha$-relaxation time $\tau_\alpha$ does depend on $k$ and on
$t_w$, and the values of $\tau_\alpha$ corresponding to each waiting
time $t_w$ are plotted in the inset (b) of the same figure. At short
times, these values are well fitted by a power law $\tau_\alpha
\approx t^z_w$ characterized by the exponent $z\approx 0.7$. In the
simulation experiment of Foffi et al. \cite{foffiaging1} this
scaling of the correlator is not fully apparent, although ``in a
crude tentative of data scaling", the authors report an exponent
$z\approx 0.38$. At this point we should mention that, beyond
detailed quantitative issues, the general predicted scenario
illustrated in Fig. \ref{fig4} is completely similar to that
reported in the simulated experiment of Foffi et al.
\cite{foffiaging1}, which, in its turn, was found to be similar to
that observed experimentally by Pham et al. \cite{pham1}.

Regarding the low-$k$ peak predicted by our theory (see Fig
\ref{fig3}), let us notice that, although the final temperature of
the quench is still above the spinodal temperature for this
isochore, the asymptotic approach of $S(k;t_{w})$ to the
non-equilibrium structure $S^{(a)}(k)$ is strongly suggestive of
some form  of \emph{arrested} spinodal decomposition. In fact,
preliminary calculations using our theory indicate that the scenario
described in the main panel and the inset of our Fig. \ref{fig3}.a
above, regarding this low-$k$ peak and the phenomenon of arrested
spinodal decomposition, is also predicted to occur at lower
concentrations, in qualitative agreement with experimental
observations (see Fig. 4.b and 4.c of Lu et al. \cite{lu1}). Further
comparisons and analysis lie, however, outside the scope of this
illustrative presentation of the possible applications of the
non-equilibrium SCGLE theory to the description of dynamic arrest
phenomena, including aging, in instantaneously quenched uniform
systems.

\section{Concluding remarks.}\label{secIV}

In this manner, in section \ref{secIII} we have illustrated with a
number of quantitative predictions for a specific model system
(involving hard sphere plus short-ranged attractive interactions)
the predictive nature of a generic theory of the non-equilibrium
irreversible evolution of the state of a homogeneous system
subjected to a homogeneous and instantaneous quenching process. This
theory is summarized by the self-consistent system of equations in
Eqs. (\ref{bdkt})-(\ref{dzdtquench}). The time-evolving state of the
system was described in terms of the static structure factor
$S(k;t_w)$ and of the $\tau$-dependence of the intermediate
scattering function $F(k,\tau; t_w)$ as a function of the waiting
time $t_w$ after the quench.

The specific process discussed  corresponds to the sudden isochoric
quench from an initial fluid state $(\phi_0,T^*_I)$ to a final state
near the ``attractive glass" transition. We observed that if the
final state is also ergodic, the structure relaxes to its value
equilibrium value $S^{eq}(k;\phi_0,T_F^*)$, whereas if the final
state is in the dynamically arrested state, the structure saturates
asymptotically to a non-equilibrium value $S^{(a)}(k;\phi_0,T_F^*)$.
In the latter case, $S(k;t_w)$ develops a non-equilibrium low-$k$
peak that indicates the appearance of spatial heterogeneities of
average size $\lambda_1 (t_{w})\approx 2\pi/k_1(t_{w})$, with $k_1$
being the position of this emerging low-$k$ maximum. The emergence
of this peak is associated with the vicinity of the gas-liquid
spinodal region. Regarding the evolution of the dynamics with aging
time, the theory predicts that the intermediate scattering function
$F(k,\tau;t_w)$ develops a two-step relaxations as the system ages.
The theory also predicts the superposition of the alpha relaxation
at different waiting times on a single master curve, well-fitted by
a stretched exponential function, as observed in the simulation and
experimental studies of aging.

Let us stress that the theory proposed in section \ref{secII},
however, is not limited to instantaneous quench processes; in
principle it is easily extendable to other quench ``programs" by
going one step back and use  Eq. (\ref{relsigmadif2pp}) instead of
Eq. (\ref{solsigmadkt}). In this manner, a number of relevant
questions could readily be addressed, such as the dependence of the
aging of $S(k;t_w)$ and $F(k,\tau; t_w)$ on the quench protocol.
Furthermore, in reality the theory of irreversible relaxation in
colloidal dispersions developed in Ref.  \cite{nescgle1}, and
summarized in Sec. \ref{secII}, is not even limited to spatially
homogeneous non-equilibrium states. The present work, however, was
meant to provide the first exploratory application of this general
theory in the simplest possible conditions. The specific results
reported here suggest that this theory provides a qualitatively and
quantitatively sound basis for the first-principles theoretical
discussion of the complex non-equilibrium phenomena associated with
the aging of structural glass-forming colloidal systems.

\bigskip

ACKNOWLEDGMENTS: We dedicate this paper to the memory of Joel
Keizer, whose theoretical conception of non-equilibrium phenomena
provided a continuous and invaluable guidance. The authors also
acknowledge Rigoberto Ju\'arez-Maldonado, Alejandro
Vizcarra-Rend\'on and Luis Enrique S\'anchez-D\'iaz for stimulating
discussions and for their continued interest in this subject. This
work was supported by the Consejo Nacional de Ciencia y
Tecnolog\'{\i}a (CONACYT, M\'{e}xico), through grants No. 84076 and
CB-2006-C01-60064, and by Fondo Mixto CONACyT-SLP through grant
FMSLP-2008-C02-107543.

\end{document}